\documentclass[a4paper,10pt]{article}

\pdfoutput=1  

\usepackage{amsmath,amssymb,mathtools}     
\usepackage{color}
\usepackage{graphicx}
\usepackage{subfigure}
\usepackage{cite}                
\usepackage{hyperref}            
\usepackage{multirow,makecell}   
\usepackage{textcomp}
\usepackage{wasysym}

\usepackage[text={17.2cm,24.5cm},centering]{geometry}

\numberwithin{equation}{section}   

\def \be {\begin{equation}}
\def \ee {\end{equation}}
\def \ba {\begin{array}}
\def \ea {\end{array}}
\def \bea{\begin{eqnarray}}
\def \eea{\end{eqnarray}}
\def \nn {\nonumber}

\def \a {\alpha}
\def \b {\beta}

\def \e {\epsilon}

\def \s {\sigma}

\def \r {\rho}

\def \mA {\mathcal A}
\def \mB {\mathcal B}

\def \mD {\mathcal D}

\def \mT {\mathcal T}

\def \p {\partial}
\def \f {\frac}

\def \mc {\mathcal}

\def \lt {\left}
\def \rt {\right}

\def \td {\tilde}

\def \lag {\langle}
\def \rag {\rangle}

\def \ep {\mathrm{e}}
\def \ii {\mathrm{i}}

\def \tr {\textrm{tr}}

\def \and {{\textrm{and}}}

\def \CFT {{\textrm{CFT}}}

\def \uum {{|0\rangle}}

\def \cl {{\textrm{cl}}}

\def \R {{\textrm{R}}}

\def \cL {{\textrm{L}}}
\def \NL {{\textrm{NL}}}
\def \NNL {{\textrm{NNL}}}
\def \NNNL {{\textrm{NNNL}}}
\def \cyl {{\textrm{cyl}}}

\begin{document}

\title{\textbf{Thermality and excited state R\'enyi entropy in two-dimensional CFT}}
\author{Feng-Li Lin$^{1}$\footnote{fengli.lin@gmail.com}~,
Huajia Wang$^{2}$\footnote{rockwhj@illinois.edu}~
and
Jia-ju Zhang$^{3,4,5}$\footnote{jiajuzhang@outlook.com}
}
\date{}

\maketitle

\vspace{-10mm}

\begin{center}
{\it
$^{1}$Department of Physics, National Taiwan Normal University, Taipei 11677, Taiwan\\\vspace{1mm}
$^{2}$Department of Physics, University of Illinois, Urbana-Champaign, IL 61801, USA\\\vspace{1mm}
$^{3}$Dipartimento di Fisica, Universit\'a degli Studi di Milano-Bicocca, Piazza della Scienza 3, I-20126 Milano, Italy\\\vspace{1mm}
$^{4}$Theoretical Physics Division, Institute of High Energy Physics, Chinese Academy of Sciences,\\19B Yuquan Rd, Beijing 100049, P.R.\,China\\\vspace{1mm}
$^{5}$Theoretical Physics Center for Science Facilities, Chinese Academy of Sciences,\\19B Yuquan Rd, Beijing 100049, P.R.\,China
}
\vspace{10mm}
\end{center}

\begin{abstract}

  We evaluate one-interval R\'enyi entropy and entanglement entropy for the excited states of two-dimensional conformal field theory (CFT) on a cylinder, and examine their differences from the ones for the thermal state.  We assume the interval to be short so that we can use operator product expansion (OPE) of twist operators to calculate R\'enyi entropy in terms of sum of one-point functions of OPE blocks.  We find that the entanglement entropy for highly excited state and thermal state behave the same way after appropriate identification of the conformal weight of the state with the temperature. However, there exists no such universal identification for the R\'enyi entropy in the short-interval expansion. Therefore, the highly excited state does not look thermal when comparing its R\'enyi entropy to the thermal state one.  As the R\'enyi entropy captures the higher moments of the reduced density matrix but the entanglement entropy only the average, our results imply that the emergence of thermality depends on how refined we look into the entanglement structure of the underlying pure excited state.

\end{abstract}

\baselineskip 18pt
\thispagestyle{empty}
\newpage

\tableofcontents


\section{Introduction}

It was conjectured that highly excited microstates behave like a thermal state, for example in the context of eigenstate thermalization hypothesis (ETH) \cite{Deutsch:1991,Srednicki:1994} when probing by few-body operators,  or canonical typicality \cite{Goldstein:2005,Popescu:2005} when considering small sub-system. This was demonstrated in two-dimensional (2D) conformal field theory (CFT) in the large central charge limit by comparing the two-point functions of light operators in the thermal state with the ones in the highly excited microstate at the leading order of large central charge expansion \cite{Fitzpatrick:2014vua,Fitzpatrick:2015zha}.  Similarly, one can also examine the conjecture in the entanglement entropy. This was done in \cite{Asplund:2014coa,Caputa:2014eta} by calculating the leading order entanglement entropy of an interval for a 2D CFT highly excited microstate on a circle, and the result agrees with the thermal entanglement entropy in \cite{Calabrese:2004eu}.
On the other hand, there is simple argument \cite{Asplund:2014coa} against extending the above agreement to all orders, which runs as follows. The entanglement entropy or R\'enyi entropy for a pure state obeys the complementarity equality, i.e., $S(A)=S(A^c)$ where $A^c$ is the complement of interval $A$. However, this equality does not hold for a mixed state such as thermal state. Thus, the full entanglement entropy or R\'enyi entropy for the highly excited state shall not equal to the ones for the thermal state. This motivates the current work to calculate the higher order results for the R\'enyi entropy of the excited states in the expansion of large central charge (i.e., corresponding to the few-body operators for general holographic CFT, which is in accordance with the requirement in the context of ETH,) and short interval (i.e., corresponding to the requirement of small sub-system in the context of canonical typicality). We then examine their differences from the thermal state R\'enyi entropy.

It is usually difficult to calculate the entanglement entropy and R\'enyi entropy for nontrivial quantum field theories, however, for a 2D CFT one can convert the problem into the calculation of correlation functions of twist operators \cite{Calabrese:2004eu}. Furthermore, we can use the operator product expansion (OPE)  to turn the product of two twist operators into a sum of the so-called OPE blocks \cite{Czech:2016xec}, i.e., the collective object of  a particular primary field and its descendants appearing in the OPE. These OPE blocks are conformal invariants. After that, one can obtain the short-interval expansion of the R\'enyi entropy by evaluating the corresponding one-point or multi-point functions of the OPE blocks. This trick has been adopted to calculate the R\'enyi entropy for some special cases \cite{Headrick:2010zt,Calabrese:2010he,Chen:2013kpa,Chen:2013dxa,Perlmutter:2013paa,Chen:2014kja,Beccaria:2014lqa,Zhang:2015hoa,Li:2016pwu,Chen:2016lbu}, such as the R\'enyi entropy of two intervals on complex plane and of the one-interval on torus.
In this paper, we adopt the same trick to investigate the short-interval expansion of excited state R\'enyi entropy on a cylinder, and then compare our results with the thermal state R\'enyi entropy.
We will consider the excited states obtained by acting on the CFT vacuum state with either primary field or non-primary fields of the vacuum family.\footnote{The states we consider are globally excited states, and they are different from the locally excited states that are investigated in, for example, \cite{He:2014mwa,Guo:2015uwa,Chen:2015usa}. One can see \cite{Sheikh-Jabbari:2016znt,Rashkov:2016xnf} for recent investigations of the globally excited state entanglement entropy using methods that are different from the method in this paper.}
In particular, we will consider the heavy state with its conformal weight comparable to the central charge because this kind of microstates are believed to behave like a thermal state, especially in the large central charge limit. We show that this is indeed the case for the entanglement entropy, which is also obtained by the other method \cite{Asplund:2014coa,Caputa:2014eta}. However, the thermal behavior will be spoiled once there are contributions of the non-vacuum OPE blocks to the excited state entanglement entropy. On the other hand, the excited state R\'enyi entropy differs from the termal state result even at the leading order, i.e. contribution from the vacuum OPE block. Our results therefore demand more detailed understanding of the thermal typicality or ETH from the R\'enyi entropy point of view.

The remaining of this paper is arranged as follows. In section~\ref{sec2} we recall a few basic facts, as well as reviewing the method of twist operator OPE, which we use to compute the short interval expansion of the R\'enyi entropy. In section~\ref{sec3} we obtain the short interval expansions of the excited state R\'enyi entropy for both heavy and light primary states. Our result shows that the excited state R\'enyi entropy for heavy states does not agree with the thermal state R\'enyi entropy. In section~\ref{sec4} we evaluate the excited state R\'enyi entropy for the non-primary states, which is found to be different from the primary one. We conclude the paper with discussion in section~\ref{sec5}. In appendix~\ref{app} we list some details of the vacuum OPE block.

\vspace{2mm}
\noindent
\textbf{Note added} When we are preparing the draft, there appears the paper \cite{Lashkari:2016vgj} in ArXiv that has some overlaps with our paper.

\section{R\'enyi entropy in 2D CFT: formulation and review}\label{sec2}

   Let the CFT be in a state $|\phi\rangle$ so that the reduced density matrix for the interval $A$ is given by
\be
\r_A=\tr_{A^c} |\phi\rag\lag\phi|.
\ee
Here $|\phi\rangle$ can be any pure state, including the vacuum state $|0\rag$. It is normalized such that $\lag\phi|\phi\rag=1$.

The R\'enyi entropy is then given by
\be
S_n= {1\over 1-n} \log \tr_A \r_A^n\;.
\ee
For $n\rightarrow 1$, it reduces to the entanglement entropy, i.e., $S_A:=S_{n\rightarrow 1}$.

By using the replica trick, the R\'enyi entropy of an interval of length $\ell$ can be obtained by the two-point function of the twist operators for the orbifold version of the original CFT, i.e.,
\be
\tr_A\r_A^n = \lag\Phi| \s(\ell)\td\s(0)  |\Phi\rag,
\ee
where $\s$ and $\td\s$ are the twist operators with conformal weights\footnote{Note that in this paper we only consider the contributions of the holomorphic sector to the R\'enyi entropy, and contributions of the anti-holomorphic sector can be calculated similarly.}
\be
h_\s=h_{\td\s}=\f{c(n^2-1)}{24n},
\ee
and the state $|\Phi\rag$ of the orbifold CFT is defined as
\be
|\Phi\rag = |\prod_{j=0}^{n-1}\phi_j\rag,
\ee
where the index $j$ labels the replicas of the original CFT.

For the ground state of a CFT on the cylinder, the one-interval R\'enyi entropy is universal, i.e., the result depends only on CFT's central charge $c$ \cite{Calabrese:2004eu}. The result has been obtained in \cite{Chen:2013kpa,Chen:2013dxa} and looks as follows (only the holomorphic part is shown)
\be\label{SnL}
S_n=\f{c(n+1)}{12n}\log \Big( \f{L}{\pi \e} \sin \f{\pi\ell}{L} \Big) =
\frac{c(n+1)}{12n} \log \f{\ell}{\e}
-\frac{ \pi^2 c (n+1) \ell^2}{72 nL^2}
-\frac{ \pi^4 c (n+1)\ell^4 }{2160 nL^4}
-\frac{ \pi^6 c (n+1) \ell^6}{34020 nL^6}
+O(\ell^8),
\ee
where $L$ is the length of the circle on which the CFT lives. In the second equality we expand the result in powers of $\ell/L$. This will be useful for later comparison.   Furthermore, by swapping the role of time and space, one can obtain the thermal state \R\'enyi entropy of an interval of length $\ell$ for a CFT on an infinite line at temperature $1/\beta$. The result is
\be\label{Snbeta}
S_n=\f{c(n+1)}{12n}\log \Big( \f{\beta}{\pi \e} \sinh \f{\pi\ell}{\b} \Big) =
\frac{c(n+1)}{12n} \log \f{\ell}{\e}
+\frac{ \pi^2 c (n+1) \ell^2}{72 n\b^2}
-\frac{ \pi^4 c (n+1)\ell^4 }{2160 n\b^4}
+\frac{ \pi^6 c (n+1) \ell^6}{34020 n\b^6}
+O(\ell^8).
\ee

\begin{figure}
\centering
\includegraphics[width=0.4\textwidth]{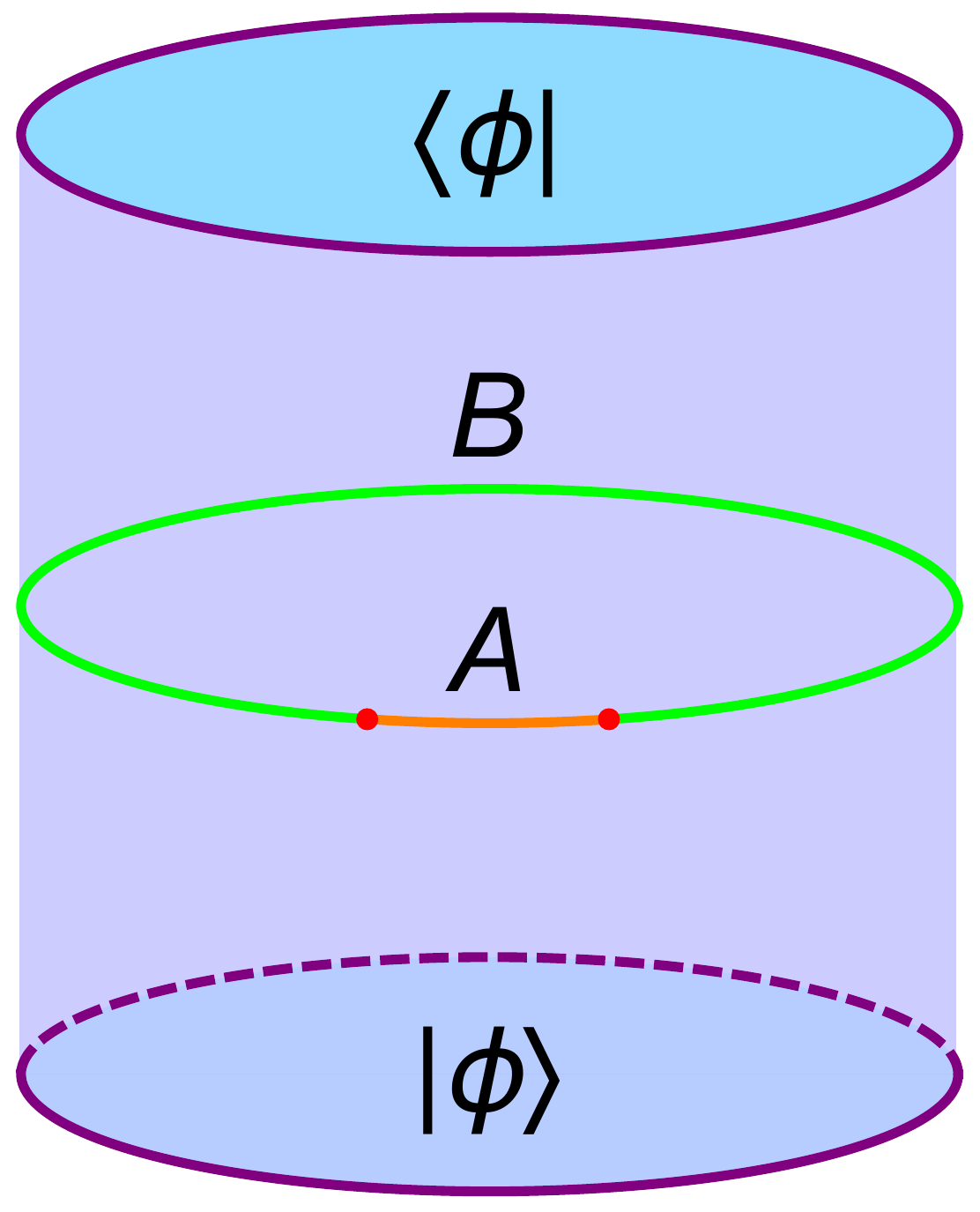}\\
\caption{The CFT setup for our calculation of the one-interval excited state R\'enyi entropy and entanglement entropy on the interval $A=[0,\ell]$. The CFT is defined on a cylinder with spatial size $L$ in the excited state $|\phi\rag$, which is created by acting on the vacuum state with an operator $\phi$.}\label{cyl}
\end{figure}

On the other hand, if $|\phi\rag$ is not the ground state,  the \R\'enyi entropy is no longer universal. This is the case considered in this paper and the setup is shown in Fig.~\ref{cyl}. The results will then depend on the details of the CFT, e.g., on the spectrum and structure constants \cite{Barrella:2013wja,Cardy:2014jwa,Chen:2014unl,Chen:2015uia,Chen:2016lbu}. In this case, we need to find some ways to obtain the approximate results.

Similar situation also holds for the CFT on tours, i.e., with a spatial circle of length $L$ and a Euclidean time circle of temperature $1/\beta$.  Moreover, the R\'enyi entropy in this case is useful for comparison with the one in the excited state of the same CFT on the cylinder.
For this purpose, it is as good to consider the results in the expansion of $\ell/L$ or $\ell/\beta$ in either low or high temperature limits. This in fact was done in \cite{Chen:2016lbu} for CFT on torus by assuming $\ell$ is small so that one can adopt the OPE of the twist operators to evaluate the one-interval R\'enyi entropy in terms of sum of one-point functions for various conformal families. The key point for this method to work is that one-point function on torus is a constant by the translational symmetry. As the same is true for the one-point function on a cylinder in excited state, we can apply the same OPE method to obtain the expansion of R\'enyi entropy.

In the following subsection, we will first describe the OPE method for evaluating the excited state R\'enyi entropy on a cylinder. Our description will be brief as the method is in parallel with the details in \cite{Chen:2016lbu} for R\'enyi entropy on torus. We will then review the results in \cite{Chen:2016lbu} for latter comparisons.

\subsection{Excited state R\'enyi entropy via OPE}

     To calculate the excited state R\'enyi entropy is just to calculate the two-point function of twist operators, i.e.,
\be\label{Renyi-1}
\tr_A \r^n_A=\lag\Phi| \s(\ell)\td\s(0)  |\Phi\rag_\cyl.
\ee

As introduced in \cite{Czech:2016xec},  the OPE of two operators of the same conformal weight can be expressed as the sum of conformal invariant objects, called the OPE blocks\footnote{If the two operators of the OPE have different conformal weights, then the OPE blocks are not conformal invariants \cite{Czech:2016xec}.  Moreover, in this paper we adopt the OPE by quasiprimary operators and their derivatives.}. Take the product of the twist operators as example, one has
\be\label{OPEb}
\s(x_1) \td\s(x_2) = |x_1-x_2|^{-2 h_{\s}} \sum_{\mathcal{O}_{\Delta}} C_{\s \td\s\mathcal{O}_{\Delta}}  B_{\Delta}(x_1,x_2)\;,
\ee
where the sum runs over all primary fields $\mathcal{O}_{\Delta}$'s, and $C_{\s \td\s\mathcal{O}_{\Delta}}$ is the OPE coefficient.  The OPE block $B_{\Delta}(x_1,x_2)$ is a succinct notation for the  sum of the primary operator $O_{\Delta}$ and its descendants appearing in the OPE. As shown in \cite{Czech:2016xec} the OPE blocks in (\ref{OPEb}) are the conformal invariant kinematic objects. Obviously, the R\'enyi entropy obtained via (\ref{Renyi-1}) and (\ref{OPEb}) will depend on the spectrum and the OPE coefficients, and thus not universal.

  It is known that the OPE block for the vacuum family is related to the modular Hamiltonian $H_A$ \cite{Czech:2016xec}. This fact leads to the first law of entanglement thermodynamics
\be
\Delta S_A= \Delta \lag H_A \rag,
\ee
where $\Delta S_A$ is the change of the entanglement entropy due to the change of CFT state $|\phi\rag$, similarly for $\Delta \lag H_A \rag$.  Beside the vacuum family, there are also contributions to the R\'enyi entropy from the other conformal family, which will then lead to the deviation from the first law.

Therefore, we can evaluate the one-point functions of the OPE blocks to obtain the R\'enyi entropy. In particular, for small $\ell$ one can express the R\'enyi entropy in the powers of $\ell$ with each term being calculated explicitly.
The contribution from the vacuum family\footnote{Note that twist operators are operators of the $n$-fold CFT, which we call $\CFT^n$. Using the vacuum family of the original CFT we can not only get the vacuum family of $\CFT^n$ but also some nonidentity conformal families. Here by vacuum family we mean that of the original CFT, not that of $\CFT^n$.} can be expanded as follows:
\bea
&& \tr_A\r_A^n=\f{c_n}{\ell^{2h_\s}}
             \bigg[ 1+b_T\lag T\rag_\phi \ell^2
                    + \Big(b_\mA\lag\mA\rag_\phi+b_{TT}\lag T\rag_\phi^2 \Big) \ell^4
                    + \Big(b_\mB\lag\mB\rag_\phi \nn\\
&&\phantom{\tr_A\r_A^ n=\f{c_n}{\ell^{2h_\s}}~}
                    + b_\mD\lag\mD\rag_\phi
                    + b_{T\mA}\lag T\rag_\phi\lag\mA\rag_\phi
                    +b_{TTT}\lag T\rag_\phi^3\Big) \ell^6 +O(\ell^8) \bigg],
\eea
with $\lag\cdots\rag_\phi := \lag\phi|\cdots|\phi\rag_\cyl$, $c_n$ is the normalization of the twist operators, and the coefficients $b$'s are given as follows\cite{Chen:2016lbu}
\bea
&& \hspace{-10mm}
   b_T=\frac{n^2-1}{12n},                                                          ~~
   b_\mA=\frac{(n^2-1)^2}{288 n^3},                                                ~~
   b_\mB=-\frac{(n^2-1)^2 \left(2 n^2(35 c+61)-93\right)}{10368 n^5(70 c+29)},     ~~
   b_\mD=\frac{(n^2-1)^3}{10368 n^5},                                              \nn\\
&& \hspace{-10mm}
   b_{TT}=\frac{(n^2-1) [5 c (n+1) (n-1)^2+2 (n^2+11)]}{1440 c n^3},               ~~
   b_{T\mA}=\frac{(n^2-1)^2 [5 c (n+1) (n-1)^2+4 (n^2+11)]}{17280 c n^5},          \nn\\
&& \hspace{-10mm}
   b_{TTT}=\frac{(n-2) (n^2-1) [35 c^2 (n+1)^2 (n-1)^3+42 c (n^2-1) (n^2+11)-16 (n+2) (n^2+47)]}
                {362880 c^2 n^5}.
\eea
We can then obtain the expansion of the excited state R\'enyi entropy contributed from the vacuum OPE block in terms of the associated one-point functions
\bea\label{REntropy}
&&\hspace{-6mm}
S^{(0)}_n= \f{c(n+1)}{12n}\log\f{\ell}{\e}   -\f{1}{n-1} \bigg\{ b_T\lag T \rag_\phi\ell^2
       +\Big[  b_\mA \lag\mA\rag_\phi
              + \big(b_{TT}-\f12 b_T^2\big)\lag T\rag_\phi^2\Big]\ell^4
       +\Big[  b_\mB \lag\mB\rag_\phi                                \nn\\
&&\hspace{-6mm}\phantom{S_n=}
          + b_\mD \lag\mD\rag_\phi
          +(b_{T\mA}-b_T b_\mA)\lag T\rag_\phi \lag\mA\rag_\phi
          + \big(b_{TTT}-b_T b_{TT}+\f13 b_T^3\big)\lag T\rag_\phi^3\Big]\ell^6
+ O(\ell^8) \bigg\}.
\eea

Furthermore, one can consider the contribution to the R\'enyi entropy from the OPE blocks other than the one of vacuum family. For the OPE block with primary field $\psi$ of conformal weight $(h_{\psi},0)$, its leading order contribution to the R\'enyi entropy up to the lowest conformal weights comes from $\psi_{j_1}\psi_{j_2}$  with $j_1<j_2$\cite{Calabrese:2010he}, and the leading order result is
\be\label{Spsi}
S^{(\psi)}_{n}= -\f{1}{n-1} b_{\psi\psi}\lag\psi\rag_\phi^2 \ell^{2h_\psi} + O(\ell^{2h_{\psi}+1}),
\ee
with
\be
b_{\psi\psi} = \f{\ii^{2h_\psi}}{\a_\psi(2n)^{2h_\psi}} \sum_{j_1<j_2} \f{1}{\big(\sin\f{(j_1-j_2))\pi}{n}\big)^{2h_\psi}}.
\ee

However, in the usual consideration of the R\'enyi entropy for the holographic CFT, it is usually assumed that there is a sparse light spectrum with no order $c$ expectation value so that the vacuum OPE block dominates the R\'enyi entropy \cite{Fitzpatrick:2015zha,Asplund:2014coa}.

\subsection{Thermal state R\'enyi entropy on a cycle}

The above formulation is in parallel with the one used in \cite{Chen:2016lbu} in which the one-interval R\'enyi entropy was calculated for CFT on a circle at finite temperature, i.e., CFT on a torus. The results in \cite{Chen:2016lbu} agree with the gravity and CFT results in \cite{Barrella:2013wja,Cardy:2014jwa,Chen:2014unl,Chen:2015uia}. The result contributed from the vacuum OPE block is the same as (\ref{REntropy}), except that the one-point functions in it should be evaluated with respect to the torus geometry, i.e., change $\lag \cdots \rag_\phi$ by $\lag\cdots\rag_\mT$ with $\mT$ denoting the torus.  As the result is useful for latter comparison with the R\'enyi entropy for highly excited state, we will write down it explicitly. Moreover, in the following we will always arrange the expansion of the result in the following manner
\be\label{powerc-exp}
S_n=S_n^\cL+S_n^\NL+S_n^\NNL+\cdots.
\ee
where $S_n$ can be either $S^{(0)}_n$ or $S^{(\psi)}_n$, and the superscript $\cL$ denotes leading order (of $1/c$ expansion), $\NL$ as next-to-leading order and so on. We will see that this is the just the large  $c$ expansion with $S_n^\cL$ of $O(c)$, $S_n^\NL$ of $O(1)$, and so on.

In the high temperature limit, i.e., $\beta \ll L$,  after evaluating the one-point functions on the torus, the resultant thermal state R\'enyi entropy is
\bea\label{Snhigh}
&& \hspace{-5mm}
   S_n^\cL = \frac{c (n+1)}{12 n} \log\f{\ell}{\e}  +\frac{\pi^2 c (n+1)}{72 n}\f{\ell^2}{\b^2}
                    + \Big( -\frac{\pi^4 c (n+1)}{2160 n}
                            -\frac{\pi^4 c (n-1) (n+1)^2 }{18 n^3}q'^2
                            +O(q'^3) \Big)\f{\ell^4}{\b^4}  \nn\\
&& \hspace{-5mm} \phantom{S_n^\cl =}
                   - \Big( -\frac{\pi^6 c (n+1)}{34020 n}
                           +\frac{\pi^6 c (n-1) (n+1)^2}{27 n^3}q'^2
                           +O(q'^3) \Big)\f{\ell^6}{\b^6} + O(\ell^8), \nn \\
&& \hspace{-5mm}
    S_{n}^\NL= - \Big( \frac{2\pi^2 (n+1)}{3 n}q'^2 + O(q'^3) \Big) \f{\ell^2}{\b^2}
               + \Big( -\frac{\pi^4 (n+1) (9 n^2-11)}{45 n^3}q'^2  +O(q'^3) \Big) \f{\ell^4}{\b^4}  \\
&& \hspace{-5mm} \phantom{S_n^\NL =}
               - \Big( \frac{2\pi^6 (n+1) (17 n^4-46 n^2+31)}{945 n^5}q'^2 + O(q'^3) \Big)\f{\ell^6}{\b^6} + O(\ell^8), \nn\\
&& \hspace{-5mm}
   S_{n}^{\NNL} =  \Big( -\frac{4\pi^4(n+1) (n^2+11)}{45n^3c}q'^4 + O(q'^5) \Big)\f{\ell^4}{\b^4}
                      -\Big( \frac{4\pi^6(n+1)(26 n^4+271 n^2-345)}{945n^5c}q'^4 + O(q'^5) \Big)\f{\ell^6}{\b^6}
                   +O(\ell^8), \nn
\eea
with $q'=\ep^{-2\pi L/\b}\ll1$. We see that the leading order result agrees with (\ref{Snbeta}) as expected, except that there are finite size corrections which are manifested in the infinite number of terms such that they are exponentially suppressed.  We can also obtain the expansion in the low temperature limit, i.e., $\beta \gg L$, and the result is related to (\ref{Snhigh}) by replacing $\b$ with $\ii L$  and $q'$ with $q:=\ep^{-2\pi\b/L}$. Again, except the infinite number of exponentially suppressed terms, the leading term agrees with (\ref{SnL}) as expected.

\section{R\'enyi entropy for a primary excited state}\label{sec3}

Given the formalism in the previous section, we can then evaluate the excited state R\'enyi entropy by calculating the one-point function of the OPE blocks.  In this section, we will consider the excited state $|\phi\rag$ by acting on the vacuum state with a holomorphic primary operator of conformal weight $h_{\phi}$. We can obtain the one-point function on the cylinder from  the one on the complex plane by the conformal transformation $z \to f(z)=\exp(2\pi \ii z/L)$.  The result for the vacuum OPE block is
\bea
&&\hspace{-5mm}
  \lag T \rag_\phi = \frac{\pi^2 (c-24 h_{\phi })}{6 L^2}, ~~
  \lag\mA \rag_\phi = \frac{\pi^4  (c (5 c+22) -240 (c+2) h_{\phi } +2880 h_{\phi }^2 )}{180 L^4}, ~~
  \lag\mB \rag_\phi = -\frac{2 \pi^6  (31 c - 504 h_{\phi } )}{525 L^6}, \nn\\ \label{onepointfunction}
&&\hspace{-5mm}
  \lag\mD \rag_\phi = \frac{\pi^6}{216 (70 c+29) L^6}\Big[  c (2 c-1) (5 c+22) (7 c+68)
                                                    -72 (70 c^3+617c^2+938c-248) h_{\phi} \\
&&\hspace{-5mm} \phantom{ \lag \mD \rag_\phi =}
                                                    +1728 (c+4) (70 c+29) h_{\phi}^2
                                                    -13824 (70 c+29) h_{\phi}^3 \Big]. \nn
\eea
Plugging these one-point functions into (\ref{REntropy}), we get the R\'enyi entropy. We now obtain  the explicit results for the state  $|\phi\rag$ to be either light or heavy, and compare them with the thermal state ones.

\subsection{The case for light state}

We first consider the state $|\phi\rag$ to be light, i.e., $h_\phi \sim 1$ so that in the large central charge limit we have $h_\phi \ll c$. We can then organize the R\'enyi entropy in the expansion of powers of central charge, i.e., as in the form of (\ref{powerc-exp}), and the results for the first few orders are
\bea \label{Snphill}
&& S_n^\cL =  \frac{c(n+1)}{12n} \log \f{\ell}{\e}
             -\frac{c \pi^2 (n+1) \ell^2}{72 n L^2}
             -\frac{c \pi^4 (n+1) \ell^4}{2160 n L^4 }
             -\frac{c \pi^6 (n+1) \ell^6}{34020 n L^6} +O(\ell^8),  \nn\\
&& S_n^\NL = \frac{\pi^2 (n+1) h_{\phi }\ell^2}{3 n L^2}
             +\frac{\pi^4 (n+1) (n^2+1) h_{\phi }\ell^4}{90 L^4 n^3}
             +\frac{2 \pi^6 (n+1) (n^4+n^2+1) h_{\phi }\ell^6}{2835 L^6 n^5} +O(\ell^8), \nn\\
&& S_n^\NNL = -\frac{\pi^4 (n+1) (n^2+11) h_{\phi }^2\ell^4}{45 c n^3 L^4}
              -\frac{\pi^6 (n+1) (2 n^4+9 n^2+37) h_{\phi }^2\ell^6}{945 c n^5 L^6} +O(\ell^8),    \\
&& S_n^\NNNL = -\frac{8 \pi^6 (n+1)(n^2-4) (n^2+47) h_{\phi }^3\ell^6}{2835 c^2 n^5 L^6}+O(\ell^8).\nn
\eea
We can take $n \to 1$ limit to obtain the entanglement entropy.

Note that the leading term agrees with (\ref{SnL}). However, there are nonzero sub-leading ${O}(1/c)$ corrections, which encode the information of the excited state as indicated by the appearance of $h_{\phi}$. Moreover, the result in (\ref{Snphill}) is also quite different from the low temperature expansion of the thermal state R\'enyi entropy, i.e., (\ref{Snhigh}) with  $\b$ replaced by $\ii L$ and $q'$ by $q:=\ep^{-2\pi\b/L}$. In the latter case, even in the leading order there are infinite number of $q$-power terms, and all the sub-leading corrections are dressed by the $q$ powers so that they vanish at zero temperature limit. These $q$ dressing terms are the non-perturbative corrections due to the finite-size effect of torus geometry in temporal direction, and they are absent for the excited state on cylinder.
This is of no wonder, since there is no correspondence between the low energy eigenstate and the low temperature thermal state. For  R\'enyi entropy there is difference at leading order, and for entanglement entropy the difference appears at the sub-leading orders. We see that R\'enyi entropy is a more powerful tool to distinguish different states.

\subsection{The case for heavy state}\label{sec32}

We now consider the case for which the excited state is heavy in large $c$ limit, i.e., $h_\phi \sim c$. In this case we introduce
\be\label{SnRE}
h_\phi=c\e_\phi,
\ee
with $\e_\phi$ being of order one. We then obtain the R\'enyi entropy contributed by the vacuum OPE block as follows
\bea
&& \hspace{-10mm}
  S^{(0)}_n =  \frac{c(n+1)}{12n} \log \f{\ell}{\e}
        +\frac{c \pi^2 (n+1) (24 \epsilon_{\phi }-1)\ell^2}{72 n L^2}
        -\frac{c \pi^4 (n+1) \Big\{ 24 \epsilon_{\phi } \big[ 2 (n^2+11) \epsilon_{\phi }- (n^2+1) \big] +n^2 \Big\} \ell^4}{2160  n^3 L^4} \nn\\
&& \hspace{-10mm} \phantom{S_n=}
-\frac{c \pi^6 (n+1) \Big\{12 \epsilon_{\phi } \big[8 (n-2) (n+2) (n^2+47) \epsilon_{\phi }^2+3 (2 n^4+9 n^2+37) \epsilon_{\phi }-2 (n^4+n^2+1)\big] + n^4 \Big\}\ell^6}{34020  n^5 L^6} \nn\\ \label{SREheavy}
&& \hspace{-10mm} \phantom{S_n=} + O(\ell^8).
\eea
We see that the result is of order $c$, and there are no sub-leading corrections. This is the result of straightforward calculation, and is also quite unexpected. The equations (3.2) happens to be not only an expansion of large central charge $c$, but also an expansion of small conformal weight $h_\phi$. After setting $h_\phi = c\epsilon_\phi$, we find that all the sub-leading terms can be absorbed into the leading term. We do not know if there is any deep reason for this result.

The result is very different from the thermal state R\'enyi entropy in the high temperature expansion, i.e., (\ref{Snhigh}). Note that the sub-leading corrections in (\ref{Snhigh}) are exponentially suppressed. In the high temperature limit, (\ref{Snhigh}) is reduced to (\ref{Snbeta}).  However, even in this limit the result is also different from (\ref{Snbeta}) because we cannot find a universal identification of $\e_\phi$ with $\beta$ to turn (\ref{SREheavy}) into (\ref{Snbeta}) term by term.

On the other hand, by taking $n\rightarrow 1$ limit we can obtain the entanglement entropy as follows
\be\label{Seee}
S^{(0)} = \frac{c}{6} \log \f{\ell}{\e}
   +\frac{c\pi^2 (24 \epsilon_{\phi }-1)\ell^2}{36 L^2}
   -\frac{c \pi^4 (24 \epsilon_{\phi }-1)^2\ell^4}{1080 L^4}
   +\frac{c\pi^6 (24 \epsilon_{\phi }-1)^3\ell^6}{17010 L^6}
   +O(\ell^8).
\ee
It is easy to see that the result (\ref{Seee}) agrees with (\ref{SnL}) of $n\rightarrow 1$ for  $\e_\phi<1/24$ if we identify $L$ in (\ref{SnL}) with $L/\sqrt{1-24\e_\phi}$. Similarly, (\ref{Seee}) agrees with (\ref{Snbeta}) of $n\rightarrow 1$ for  $\e_\phi>1/24$ if we identify $\b$ in (\ref{Snbeta}) with $L/\sqrt{24\e_\phi-1}$.
This implies that the excited state entanglement entropy behaves the same way as the ground/thermal state entanglement entropy in the low/high temperature limit. This is consistent with the expectation of ETH \cite{Deutsch:1991,Srednicki:1994} or canonical typicality \cite{Goldstein:2005,Popescu:2005}, i.e., a typical eigenstate can mimic the thermal or finite-size effect. Moreover, for $\e_\phi=1/24$ we have
\be
S^{(0)} = \frac{c}{6} \log \f{\ell}{\e},
\ee
which behaves like the ground state entanglement entropy on a complex plane.

\begin{figure}
\centering
\includegraphics[width=\textwidth]{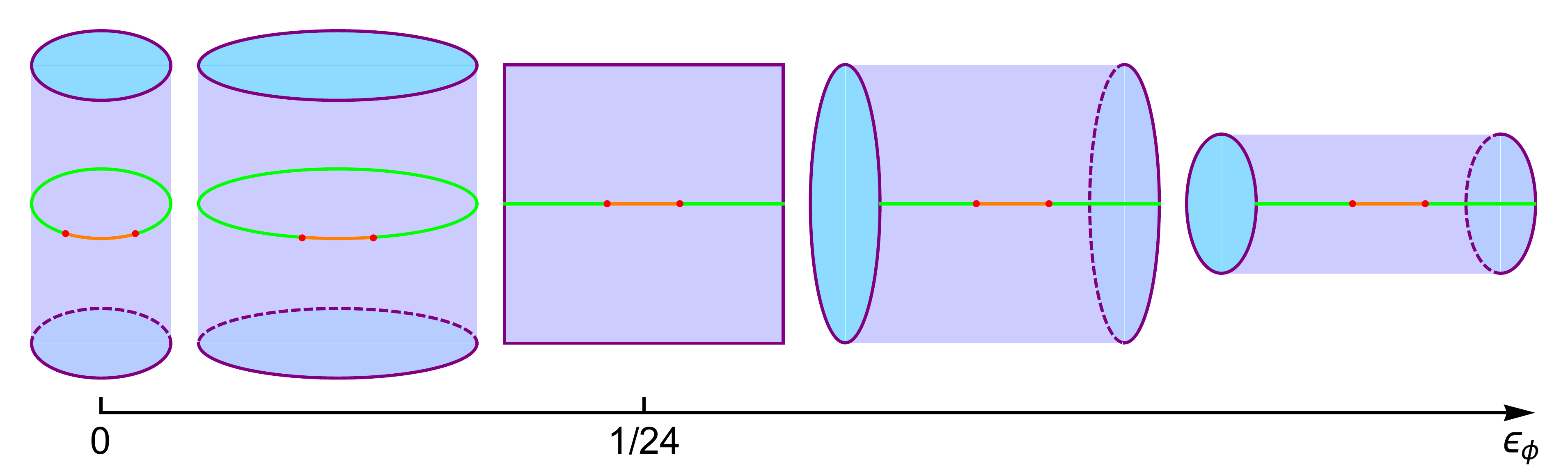}\\
\caption{When we tune the parameter $\e_\phi$ from $\e_\phi<1/24$ to $\e_\phi>1/24$, the entanglement entropy of the short interval on cylinder in excited state $|\phi\rag$ with $h_\phi=c h_\phi$ behaves as the entanglement entropies of the same interval length on different manifolds. When $\e_\phi<1/24$ there are cylinders with periodic boundary condition in the spatial direction, and the cylinder becomes `fatter' when $\e_\phi$ becomes larger. When $\e_\phi=0$, it is the complex plane. When $\e_\phi>1/24$ there are cylinders with periodic boundary condition in the temporal direction, and the cylinder becomes `thinner' when $\e_\phi$ becomes larger.  }\label{identification}
\end{figure}

In summary, the one-interval entanglement entropy with interval length $\ell$ for a heavy pure state $|\phi\rangle$ with $h_\phi=c\e_\phi$ on a circle of length $L$ is equivalent to the one-interval entanglement entropy with the same interval length for the CFT ground state or thermal state depending on the value of $\e_\phi$ as following. (i) When $\e_\phi<1/24$ it behaves as the entanglement entropy for the CFT ground state on a circle of length $L/\sqrt{1-24\e_\phi};$\footnote{We may also say that it behaves as the entanglement entropy of an interval of length $\ell\sqrt{1-24\e_\phi}$ for a CFT living on a circle of length $L$. Since we are discussing about a CFT, the two statements are in fact equivalent.} (ii) when $\e_\phi>1/24$ it behaves as the entanglement entropy for a CFT thermal state on an infinite straight line with temperature $\sqrt{24\e_\phi-1}/L$; and (iii) when $\e_\phi=1/24$ it behaves as the entanglement entropy for the CFT in ground state on an infinite straight line. We summarize the results in Fig.~\ref{identification} by tuning the parameter $\e_\phi$ from $\e_\phi<1/24$ to $\e_\phi>1/24$.

Through just straightforward calculation, our result here is quite provoking. In summary, the high temperature limit is the same as the micro-canoical emsemble, and thus our result can be phrased as follows: both the excited state R\'enyi entropy and the entanglement entropy are different from the thermal state ones of the canonical ensemble by the sub-leading corrections in the large $c$ expansion.
However, in the high temperature limit, i.e., the canonical ensemble is turned into the micro-canonical one, the excited state entanglement entropy agrees with the thermal state one. This agrees with what one will expect from ETH or canonical typicality. On the other hand, the excited R\'enyi entropy still fails to behave as the micro-canonical thermal state one because without taking $n \rightarrow 1$ limit it is impossible to identify a universal temperature in terms of $\epsilon_{\phi}$  for all orders of $\ell$ expansion of (\ref{SREheavy}). The R\'enyi entropy encodes higher moments of the reduced density matrix than the entanglement entropy, our above result implies that the emergence of thermality of the excited state depends on how detailed we compare the entanglement structures.

\subsection{Leading contribution from non-vacuum OPE block}

   By following the discussion around (\ref{Spsi}), we now evaluate the leading order result of the excited state R\'enyi entropy contributed from the non-vacuum OPE block. For a holomorphic primary operator $\psi$ of conformal weight $h_\psi$, we have the one-point function as following
\be
\lag\psi\rag_\phi=\Big( \f{2\pi\ii}{L} \Big)^{h_\psi}C_{\phi\psi\phi}.
\ee
where $C_{\phi\psi\phi}$ is the OPE coefficient. Therefore, there is nonzero contribution to the one-interval R\'enyi entropy for the excited state $|\phi\rag$ from the corresponding OPE block characterized by the primary field $\psi$ only if the structure constant $C_{\phi\psi\phi}$ is nonvanishing. In this case from (\ref{Spsi}) we have
\be
S^{(\psi)}_n \sim\f{C_{\phi\psi\phi}^2}{\a_\psi} \Big( \f{\ell}{L} \Big)^{2h_\psi} + O(\ell^{2h_{\psi}+1}).
\ee
As the result depends on the OPE coefficient, it is not universal. Moreover, by taking $n\rightarrow 1$ limit we obtain the nonzero contribution to the entanglement entropy, which will then spoil the thermal behavior of (\ref{Seee}) for the vacuum OPE block.  On the other hand, if there is a gap in the low energy spectrum of CFT as it is usually assumed for the holographic CFT, then the non-thermal deviation from the non-vacuum OPE blocks will be highly suppressed.

\subsection{Check ETH}
   If ETH holds, one would expect that the one-point functions of some set of local physical operators $\{ \mathcal{O}_{phy} \}$ for some microstate are the same as the ones for the thermal state. In the context of this paper, we expect
\be \label{OphiETH}
\lag \mathcal{O}_{phy} \rag_\phi = \lag \mathcal{O}_{phy} \rag_\mT
\ee
after the following identification of the conformal weight $h_{\phi}$ and the inverse temperature $\beta$:
\be \label{betaandhphi}
h_{\phi}=c \epsilon_{\phi}, ~~ \b=\f{L}{\sqrt{24\e_\phi-1}}.
\ee
In section~\ref{sec32} we have shown that the excited state R\'enyi entropy does not match with the thermal state one even after the identification of (\ref{betaandhphi}), despite that the entanglement entropy does match.  We now like to check if the ETH relation holds universally for all physical observables or not.

We first consider the case when $\mathcal{O}_{phy}=T$. The result of $\lag T \rag_{\phi}$ is given in (\ref{onepointfunction}), and the result for $ \lag T \rag_\mT$ in high temperature is given in \cite{Chen:2016lbu}, i.e., it is
\be \label{ThighT}
 \hspace{-10mm}
   \lag T \rag_\mT = -\frac{\pi^2 c}{6 \beta^2}
                     +\frac{8 \pi^2}{\beta^2}q'^2
                     +\frac{12 \pi^2}{\beta^2}q'^3
                     +\frac{24 \pi^2}{\beta^2}q'^4
                     +O(q'^5),
\ee
with  $q':=\ep^{-2\pi L/\b}$. If the relation (\ref{OphiETH}) holds for this case, then we get the following identification
\be \label{hphiandbeta}
h_{\phi} = \frac{c (L^2+\beta^2)}{24 \beta^2}
          -\frac{2 L^2 q'^2}{\beta^2}
          -\frac{3 L^2 q'^3}{\beta2}
          -\frac{6 L^2 q'^4}{\beta^2}
          +O(q'^5).
\ee
Note that in the large $c$ limit, it reproduces (\ref{betaandhphi}).  We now substitute (\ref{hphiandbeta}) into (\ref{onepointfunction}) to obtain
\be \label{AphiETH}
\lag \mA \rag_\phi =  \frac{\pi^4 c (5 c L^4-20 L^2 \beta^2+2 \beta^4)}{180 L^4 \beta^4}
                     -\frac{8 \pi^4 (c L^2-2 \beta^2)q'^2}{3 L^2 \beta^4}
                     -\frac{4 \pi^4 (c L^2-2 \beta^2)q'^3}{L^2 \beta^4}
                     -\frac{8 \pi^4 \big( (c-8) L^2-2 \beta^2 \big)q'^4}{L^2 \beta^4}
                     + O(q'^5),
\ee
which is very different from the thermal state one given in \cite{Chen:2016lbu}, i.e.,
\bea \label{AhighT}
\hspace{-10mm}
   \lag \mA \rag_\mT = \frac{\pi^4 c (5 c+22)}{180 \beta^4}
                      +\frac{8 \pi^4 (5 c+22)}{3 \beta^4}q'^2
                      +\frac{12 \pi^4 (5 c+22)}{\beta^4}q'^3
                      +\frac{168 \pi^4 (c+6)}{\beta^4}q'^4
                      +O(q^5). \nn
\eea
Therefore, the ETH relation (\ref{OphiETH}) does not hold universally for the descendant operators of the vacuum family.

From the above, we can also infer that the relation (\ref{OphiETH}) will not hold universally for the other primary field $\mathcal{O}$ because $\lag \mathcal{O} \rag_\mT$ depends on all the non-vanishing structure constants $C_{\varphi\mathcal{O}\varphi}$ of $\mathcal{O}$ with all primary operators $\{ \varphi \}$, but $\lag \mathcal{O} \rag_\phi$ only depend on the structure constant $C_{\phi\mathcal{O}\phi}$. For example, if $C_{\phi\mathcal{O}\phi}=0$, we have $\lag \mathcal{O} \rag_\phi=0$, but generally $\lag \mathcal{O} \rag_\mT \neq 0$.

\section{R\'enyi entropy for descendant states of vacuum family}\label{sec4}

  For curiosity, we now consider the R\'enyi entropy for the excited state $|\phi\rag$ obtained by acting on the vacuum state with the descendant operator of the vacuum family, i.e., explicitly we will consider the following $|\phi\rag$,
\be
|\phi\rag =  \Big\{ \uum,
                    \f{1}{\sqrt{\a_T}} |T \rag,
                    \f{1}{\sqrt{\a_{\p T}}} |\p T \rag,
                    \f{1}{\sqrt{\a_{\p^2 T}}} |\p^2 T \rag,
                    \f{1}{\sqrt{\a_\mA}} |\mA \rag \Big\},
\ee
with
\be
\a_T=\frac{c}{2}, ~~ \a_{\p T}=2c, ~~ \a_{\p^2 T}=20c, ~~ \a_\mA=\frac{c(5c+22)}{10}.
\ee
For such kind of excited states, we only need to consider the contribution to the R\'enyi entropy from the vacuum OPE block because the one-point function of the non-vacuum OPE block in this kind of state is zero, i.e.,
\be
\lag B_{\Delta\ne 0} \rag_\phi =0.
\ee

  The first thing we need for carrying out the evaluation is the one-point function of the operators in vacuum OPE block. The result is as follows
\bea
&& \lag T \rag_\phi = \Big\{ \frac{\pi ^2 c}{6 L^2},
                             \frac{\pi ^2 (c-48)}{6 L^2},
                             \frac{\pi ^2 (c-72)}{6 L^2},
                             \frac{\pi ^2 (c-96)}{6 L^2},
                             \frac{\pi ^2 (c-96)}{6 L^2} \Big\},                      \nn\\
&& \lag \mA \rag_\phi = \Big\{ \frac{\pi ^4 c (5 c+22)}{180 L^4},
                               \frac{\pi ^4 (c+480) (5 c+22)}{180 L^4},
                               \frac{\pi ^4 (c+2160) (5 c+22)}{180 L^4},              \nn\\
&& \phantom{\lag \mA \rag_\phi = \Big\{}
                               \frac{\pi ^4 (c+5568) (5 c+22)}{180 L^4},
                               \frac{\pi ^4 (5 c^2+4822c+69504)}{180 L^4} \Big\},    \nn\\
&& \lag \mB \rag_\phi = \Big\{ -\frac{62 \pi ^6 c}{525 L^6},
                                \frac{2 \pi ^6 (120929 c+1008)}{525 L^6},
                                \frac{2 \pi ^6 (1088609 c+243432)}{525 L^6},          \nn\\
&& \phantom{\lag \mB \rag_\phi = \Big\{}
                                \frac{2 \pi ^6 (4838369 c+1477728)}{525 L^6},
                                \frac{2 \pi ^6 (241889 c+1066464)}{525 L^6} \Big\},                             \\
&& \lag \mD \rag_\phi = \Big\{ \frac{\pi ^6 c (2 c-1) (5 c+22) (7 c+68)}{216 (70 c+29) L^6},
                               \frac{\pi ^6 (c+1584) (2 c-1) (5 c+22) (7 c+68)}{216 (70 c+29) L^6},             \nn\\
&& \phantom{\lag \mD \rag_\phi = \Big\{}
                               \frac{\pi ^6 (c+6696) (2 c-1) (5 c+22) (7 c+68)}{216 (70 c+29) L^6},
                               \frac{\pi ^6 (c+16992) (2 c-1) (5 c+22) (7 c+68)}{216 (70 c+29) L^6},            \nn\\
&& \phantom{\lag \mD \rag_\phi = \Big\{}
                               \frac{\pi ^6 (2 c-1) (7 c+68) (5 c^2+15862c-614592)}{216 (70 c+29) L^6} \Big\}. \nn
\eea
We also include the results for the vacuum state in the first entries for comparison, and its R\'enyi entropy has been obtained in this way in \cite{Chen:2013kpa,Chen:2013dxa}. It is nothing but  (\ref{SnL}).

For   $|\phi\rag=\f{1}{\sqrt{\a_T}}|T\rag$, we obtain the corresponding R\'enyi entropy
\bea
&& S_n^\cL =  \f{c(n+1)}{12n}\log\f{\ell}{\e}
             -\frac{\pi ^2 c (n+1) \ell ^2}{72 n L^2 }
             -\frac{\pi ^4 c (n+1) (121 n^2-120) \ell ^4}{2160 n^3 L^4 }  \nn\\
&& \phantom{S_n^\cL =}
             -\frac{\pi ^6 c (n+1) (n^4-1260 n^2+1260) \ell ^6}{34020 n^5 L^6 }
             +O(\ell ^8),                                                             \nn\\
&& S_n^\NL = \frac{2\pi ^2 (n+1) \ell ^2}{3 L^2 n}
            +\frac{\ell ^4(n+1) (n^2+1) \pi ^4}{45 L^4 n^3}
            +\frac{4 \pi ^6(n+1) (22 n^4+211 n^2-230) \ell ^6}{2835 L^6 n^5}
            +O(\ell ^8),                                                              \nn\\
&& S_n^\NNL = -\frac{4\pi ^4 (n+1) (n^2+11) \ell ^4}{45 c n^3 L^4 }
              -\frac{4\pi ^6 (n+1) (2 n^4+9 n^2+37) \ell ^6}{945 c n^5 L^6 }
              +O(\ell ^8),                                                            \\
&& S_n^\NNNL = -\frac{64 \pi ^6(n+1)(n^2-4) (n^2+47) \ell ^6}{2835 c^2 n^5 L^6}
               +O(\ell ^8).                                                           \nn
\eea
As expected, the result is different from the (\ref{Snphill}) with $h_\phi=2$ for the primary excited state. More importantly, it is interesting to see that the difference even occurs at the leading order, i.e., $S_n^\cL$ is different from (\ref{SnL}).  This is in contrast to the case for the entanglement entropy by taking $n\rightarrow 1$, and the result is
\bea
&& S^\cL =  \frac{c}{6} \log \f{\ell}{\e}
            -\frac{\pi ^2 c \ell ^2}{36 L^2}
            -\frac{\pi ^4 c \ell ^4}{1080 L^4}
            -\frac{\pi ^6 c \ell ^6}{17010 L^6}
             +O(\ell^8),  \nn\\
&& S^\NL = \frac{4 \pi ^2 \ell ^2}{3 L^2}
          +\frac{4 \pi ^4 \ell ^4}{45 L^4}
          +\frac{8 \pi ^6 \ell ^6}{945 L^6}
          +O(\ell ^8) , \nn\\
&& S^\NNL = -\frac{32 \pi ^4 \ell ^4}{15 c L^4}
            -\frac{128 \pi ^6 \ell ^6}{315 c L^6}
            +O(\ell ^8) , \\
&& S^\NNNL = \frac{2048 \pi ^6 \ell ^6}{315 c^2 L^6}
           +O(\ell ^8). \nn
\eea
Note that this is the same as (\ref{Snphill}) of $n\rightarrow 1$ by setting $h_\phi=2$. Thus, we have checked up to order $\ell^6$ that with the contributions only from the vacuum family, the entanglement entropy cannot distinguish the primary state from the vacuum descendant state with the same conformal weight; on the other hand, the R\'enyi entropy can.

There are similar stories for the other three excited states, and we just list the R\'enyi entropies. For the case with $|\phi\rag=\f{1}{\sqrt{\a_{\p T}}}|\p T\rag$, we have
\bea
&& S_n^\cL = \f{c(n+1)}{12n}\log\f{\ell}{\e}
            -\frac{\pi ^2 c (n+1) \ell ^2}{72 n L^2}
            -\frac{\pi ^4 c (n+1) (481 n^2-480) \ell ^4}{2160 n^3 L^4 }  \nn\\
&& \phantom{S_n^\cL =}
            +\frac{\pi ^6 c (n+1) (5039 n^4+2520 n^2-7560) \ell ^6}{34020 L^6 n^5}
             +O(\ell ^8),                                                             \nn\\
&& S_n^\NL =
            \frac{\pi ^2(n+1) \ell ^2}{ n L^2}
            -\frac{\pi ^4(n+1) (37 n^2-43) \ell ^4}{90 n^3 L^4}
            +\frac{2 \pi ^6(n+1) (57 n^4+911 n^2-965) \ell ^6}{945 n^5 L^6}+O(\ell ^8),                                                              \nn\\
&& S_n^\NNL =
              -\frac{\pi ^4(n+1) \left(n^2+11\right) \ell ^4}{5 c n^3 L^4}
              +\frac{\pi ^6(n+1) (106 n^4+1093 n^2-1343) \ell ^6}{315 c n^5 L^6}+O(\ell ^8),                                                            \\
&& S_n^\NNNL =
               -\frac{8 \pi ^6(n+1) (n^2-4) (n^2+47) \ell ^6}{105 c^2 n^5 L^6 }+O(\ell ^8).                                                           \nn
\eea

For the case with $|\phi\rag=\f{1}{\sqrt{\a_{\p^2 T}}}|\p^2 T\rag$, we have
\bea
&& S_n^\cL =  \f{c(n+1)}{12n}\log\f{\ell}{\e}
             -\frac{\pi ^2 c (n+1) \ell ^2}{72 n L^2}
             -\frac{\pi ^4 c (n+1) (1201 n^2-1200) \ell ^4}{2160 n^3 L^4}  \nn\\
&& \phantom{S_n^\cL =}
             +\frac{\pi ^6 c (n+1) (32759 n^4-7560 n^2-25200) \ell ^6}{34020 n^5 L^6}
             +O(\ell ^8),                                                             \nn\\
&& S_n^\NL =
            \frac{4 (n+1) \pi ^2 \ell ^2}{3 L^2 n}
            -\frac{2 \pi ^4(n+1) (33 n^2-35) \ell ^4}{45 n^3 L^4}
            +\frac{2 \pi ^6(n+1) (1264 n^4+8047 n^2-9299) \ell ^6}{2835 n^5 L^6}
            +O(\ell ^8),                                                              \nn\\
&& S_n^\NNL =
              -\frac{16\pi ^4 (n+1) (n^2+11) \ell ^4}{45 c n^3 L^4 }
              +\frac{16 \pi ^6 (n+1) (466 n^4+4715 n^2-5421) \ell ^6}{4725 c n^5 L^6}
              +O(\ell ^8),                                                            \\
&& S_n^\NNNL =
               -\frac{512 \pi ^6(n+1)(n^2-4) (n^2+47) \ell ^6}{2835 c^2 n^5 L^6}
               +O(\ell ^8).                                                           \nn
\eea

For the case with $|\phi\rag=\f{1}{\sqrt{\a_{\mA}}}|\mA\rag$, we have
\bea
&& S_n^\cL =  \f{c(n+1)}{12n}\log\f{\ell}{\e}
             -\frac{\pi ^2 c (n+1) \ell ^2}{72 n L^2}
             -\frac{\pi ^4 c (n+1) (241 n^2-240)\ell ^4}{2160 n^3 L^4}  \nn\\
&& \phantom{S_n^\cL =}
             -\frac{\pi ^6 c (n+1) (n^4-2520 n^2+2520) \ell ^6}{34020 n^5 L^6}
             +O(\ell ^8),                                                             \nn\\
&& S_n^\NL =
            \frac{4 \pi ^2(n+1) \ell ^2}{3 n L^2}
            -\frac{4 \pi ^4 (n+1) (5 n^2-6) \ell ^4}{45 n^3 L^4}
            +\frac{4 \pi ^6(n+1) (86 n^4+1073 n^2-1153) \ell ^6}{2835 n^5 L^6}
            +O(\ell ^8),                                                              \nn\\
&& S_n^\NNL =
              -\frac{16\pi ^4 (n+1) (n^2+11) \ell ^4}{45 c n^3 L^4}
              +\frac{16\pi ^6 (n+1) (144 n^4+1495 n^2-1879) \ell ^6}{4725 c n^5 L^6}
              +O(\ell ^8),                                                            \\
&& S_n^\NNNL =
               -\frac{512\pi ^6 (n+1)(n^2-4) (n^2+47) \ell ^6}{2835 c^2 n^5 L^6}
               +O(\ell ^8).                                                           \nn
\eea

\section{Conclusion and discussion}\label{sec5}

In this paper we calculate the large central charge and short-interval expansion of the one-interval R\'enyi entropy and entanglement entropy for the excited state of a 2D CFT on a circle. Our primary goal is to compare the result with the thermal state R\'enyi entropy and entanglement entropy, and see if these entanglement quantities can tell a highly exited state from the thermal state or not.  To carry out the calculation we adopt the trick of OPE to turn the two-point function of the twist operators in the excited state of the replica CFT, which is the replica of the reduced density matrix for evaluating R\'enyi entropy,  into sum of the one point functions of the OPE blocks. After evaluating the associated one-point function we can then obtain the R\'enyi entropy and entanglement entropy in the short-interval expansion up to the sixth order for both heavy (i.e., with its conformal weight order of the central charge) and light (i.e., with order one conformal weight) pure states.

We then compare our results with the previous results for the thermal state R\'enyi entropy and entanglement entropy.  We first consider the contribution to the R\'enyi entropy only from the vacuum OPE block. Our result shows that unlike the thermal state case, the excited state R\'enyi entropy or entanglement entropy for the heavy state receive no sub-leading correction in the large central charge expansion. The absence of such corrections indicates that thermality apparently fails for both the R\'enyi entropy and entanglement entropy. However, in the high temperature limit, we can neglect these exponentially suppressed corrections. We then find that the short-interval epansion for the excited state entanglement entropy agrees with the thermal state expansion after identifying an effective temperature corresponding to the heavy state. This agrees with the expectation from the ETH or canonical typicality, as well as the earlier checks in the literatures. However, when the R\'enyi entropy is considered, we find it impossible to identify such an effective temperature that make the short-interval expansion of the excited state R\'enyi entropy agree with the thermal state one.

Though our result is obtained straightforwardly, it is striking in the sense that we explicitly demonstrate that the thermality of the heavy pure state fails for R\'enyi entropy, while it holds for entanglement entropy. In some sense, the R\'enyi entropy encodes the higher moments of the reduced density matrix, our results implies that the thermality of heavy pure state emerges in the entanglement structure only in the average sense, i.e., the entanglement entropy, but not in the more refined entanglement structure, i.e., the R\'enyi entropy. If our observation holds in the more general cases such as beyond the context of CFT, then it is the caveat when one tries to formulate or apply the ETH or canonical typicality.

Besides, we also consider the contribution to the excited state R\'enyi entropy and entanglement entropy from the non-vacuum OPE blocks, which cannot be neglected if there is no low energy gap in the CFT spectrum. In this case, it is easy to see that the thermality of the heavy pure state fails even for the entanglement entropy.  However, there are evidences showing that the holographic CFT is gapped so that this kind of corrections can be neglected in the large central charge limit. Finally, we also consider the excited state created by acting on CFT's vacuum state  with the descendant operators of the vacuum family.

To summarize, we find in this paper that the whether thermality emerges or not depends on how refined we look into the entanglement structure of the underlying pure state. The calculations in this paper provide some clues for further studies. In this paper we examine the issue by the short-interval expansion of the R\'enyi entropy, and there may have a rare chance that the thermality may be encoded in a highly nontrivial way in the full R\'enyi entropy. In any case, our results could be a step stone to the route of uncovering the caveats for the emergence of thermality beyond the context of ETH and canonical typicality.

\section*{Acknowledgments}

We would like to thank Bin Chen, Thomas Faulkner, Song He and Jie-qiang Wu for valuable discussions.
JJZ would like to thank Peking University for hospitality during the concluding stage of this work.
 We thank Matthew Headrick for his Mathematica code \emph{Virasoro.nb} that could be downloaded at \url{http://people.brandeis.edu/~headrick/Mathematica/index.html}.
FLL was supported by Taiwan Ministry of Science and Technology through Grant No.~103-2112-M-003-001-MY3 and No.~103-2811-M-003-024.
HJW was supported by DARPA YFA contract D15AP00108.
JJZ was supported by NSFC Grant No.~11222549 and No.~11575202.

\appendix

\section{Some details of vacuum OPE block}\label{app}

We list the holomorphic quasiprimary operators in vacuum family to level 6.
In level 2, we has the quasiprimary operator $T$, with the usual normalization $\a_T=\f{c}{2}$. In level 4, we have
\be
\mA=(TT)-\f{3}{10}\p^2T, ~~ \a_{\mc A}=\f{c(5c+22)}{10}.
\ee
In level 6, we have
\bea
&& \mB=(\p T\p T)-\f{4}{5}(\p^2TT)-\f{1}{42}\p^4T, ~~
    \a_{\mc B}=\frac{36c (70 c+29)}{175},                                    \nn\\
&& \mD=(T(TT))-\f{9}{10}(\p^2TT)-\f{1}{28}\p^4 T +\f{93}{70c+29} \mc B,      \\
&& \a_{\mc D}=\frac{3 c (2 c-1) (5 c+22) (7 c+68)}{4 (70 c+29)}.             \nn
\eea
Under a general coordinate transformation $z\to f(z)$, we have
\bea
&& T(z)=f'^2 T(f)+\f{c}{12}s, ~~
   \mA(z)=f'^4\mA(f)+\f{5c+22}{30}s \Big( f'^2 T(f)+\f{c}{24}s \Big),                     \nn\\
&& \mB(z)= f'^6\mB(f)-\f{8}{5}f'^4s\mA(f)
          -\f{70c+29}{1050}f'^4s\p^2T(f)
          +\f{70c+29}{420}f'^2(f's'-2f''s)\p T(f)                                         \nn\\
&& \phantom{\mB(z)=}
          -\f{1}{1050}\lt( 28(5c+22)f'^2s^2+(70c+29)(f'^2s''-5f'f''s'+5f''^2s) \rt)T(f)   \\
&& \phantom{\mB(z)=}
          -\f{c}{50400}\lt( 744s^3+ (70c+29)(4ss''-5s'^2) \rt),                           \nn\\
&& \mD(z)=f'^6\mD(f)+\f{(2c-1)(7c+68)}{70c+29}s \Big( \f{5}{4} f'^4\mA(f)+\f{5c+22}{48}s \big( f'^2T(f)+\f{c}{36}s \big) \Big),   \nn
\eea
with the definition of Schwarzian derivative
\be
s(z)=\f{f'''(z)}{f'(z)}-\f32 \bigg( \f{f''(z)}{f'(z)} \bigg)^2.
\ee
In the calculations we need the structure constants
\bea
&& C_{TTT}=c, ~~
   C_{TT\mA}=\frac{c(5c+22)}{10}, ~~
   C_{TT\mB}=-\frac{2c(70 c+29)}{35},                           \nn\\
&& C_{TT\mD}=0, ~~
   C_{T\mA\mA}=\frac{2 c (5 c+22)}{5}, ~~
   C_{\mA\mA\mA}=\frac{c (5 c+22) (5 c+64)}{25},                \\
&& C_{\mA\mA\mB}=-\frac{4 c (5 c+22) (14 c+73)}{35}, ~~
   C_{\mA\mA\mD}=\frac{6 c (2 c-1) (5 c+22) (7 c+68)}{70 c+29}. \nn
\eea
For a general primary operator $\phi$ with conformal weight $h_\phi$ and normalization factor $\a_\phi=1$, we have the structure constants
\bea
&& C_{\phi T \phi}=h_\phi, ~~~
   C_{\phi\mA\phi}=\f{h_\phi(5h_\phi+1)}{5}, ~~~
   C_{\phi\mB\phi}= -\frac{2 h_{\phi}(14 h_{\phi}+1)}{35} ,  \nn\\
&& C_{\phi\mD\phi}= \frac{h_{\phi} [(70 c+29) h_{\phi}^2+(42 c-57) h_{\phi}+(8 c-2)] }{70 c+29}.
\eea


\providecommand{\href}[2]{#2}\begingroup\raggedright\endgroup

\end{document}